\documentclass{article}

\usepackage{amssymb}
\usepackage{spconf,amsmath,graphicx}
\usepackage{mathtools}
\usepackage{lipsum}
\usepackage{upgreek}
\usepackage{dirtytalk}
\usepackage{siunitx}
\usepackage{mathrsfs}
\usepackage{float}
\usepackage[caption=false,font=footnotesize]{subfig}
\usepackage{dirtytalk}
\usepackage{hyperref}
\usepackage[dvipsnames]{xcolor}
\usepackage{pifont}
\usepackage{bm,upgreek}
\usepackage{anyfontsize}
\usepackage[backend=biber,style=ieee,giveninits=true,sorting=none,doi=false,url=false,isbn=false,eprint=false,maxbibnames=99]{biblatex}
\usepackage{microtype}
\usepackage{enumitem}

\makeatletter
\newcommand{\mathleft}{\@fleqntrue\@mathmargin0pt}
\newcommand{\mathcenter}{\@fleqnfalse}
\makeatother

\hypersetup{
    pdffitwindow=true,
    colorlinks=true,
    linkcolor={blue},
    citecolor={blue},
    urlcolor={black}
}

\allowdisplaybreaks

\title{EFFECT OF TARGET SIGNALS AND DELAYS ON SPATIALLY SELECTIVE \\ ACTIVE NOISE CONTROL FOR OPEN-FITTING HEARABLES}
\name{Tong Xiao, Simon Doclo\thanks{This research was funded by the Deutsche Forschungsgemeinschaft (DFG, German Research Foundation) -- Project-ID 352015383 -- SFB 1330 C1.}}
\address{Department of Medical Physics and Acoustics and Cluster of Excellence Hearing4all, \\ Carl von Ossietzky Universit\"{a}t Oldenburg, Germany \\[1mm]
\href{mailto:tong.xiao@uni-oldenburg.de}{tong.xiao@uni-oldenburg.de} $\qquad$
\href{mailto:simon.doclo@uni-oldenburg.de}{simon.doclo@uni-oldenburg.de}
}

\begin{document}
\ninept
\maketitle
\begin{abstract}
Spatially selective active noise control (ANC) hearables are designed to reduce unwanted noise from certain directions while preserving desired sounds from other directions. In previous studies, the target signal has been defined either as the delayed desired component in one of the reference microphone signals or as the desired component in the error microphone signal without any delay. In this paper, we systematically investigate the influence of delays in different target signals on the ANC performance and provide an intuitive explanation for how the system obtains the desired signal. Simulations were conducted on a pair of open-fitting hearables for localized speech and noise sources in an anechoic environment. The performance was assessed in terms of noise reduction, signal quality and control effort. Results indicate that optimal performance is achieved without delays when the target signal is defined at the error microphone, whereas causality necessitates delays when the target signal is defined at the reference microphone. The optimal delay is found to be the acoustic delay between this reference microphone and the error microphone from the desired source.

\end{abstract}


\section{Introduction}
\label{sec:intro}
Active noise control (ANC) hearables are designed to create a quiet environment by using secondary sources to generate anti-noise, aiming at minimizing sound at certain positions when superimposed on the primary noise~\cite{Elliott2000,Hansen2012}. Based on their fit, hearables can be categorized as closed-fitting (completely occluding the ear), open-fitting (partially occluding the ear), and open-ear (no occlusion). Recent research focuses on designing intelligent ANC hearables with spatial selectivity, especially for complex acoustic environments like cocktail-party scenarios with multiple sound sources from different directions~\cite{kajikawa2012recent,Chang2016Listening,gupta2022augmented}. In these environments, users may want to focus on desired sounds from a specific direction (e.g., from the front) while blocking out undesired sounds from other directions.

Modern ANC hearables are commonly equipped with multiple microphones, including both reference microphones on the exterior of the hearable and error microphones in the interior close to the ear canal. Hence, beamforming can be used to enhance a sound source from a certain direction and reduce sound sources from other directions~\cite{vanveen1988,Doclo2015,gannot2017consolidated}, e.g., using the linearly-constrained-minimum-power (LCMP) or the minimum-power-distortionless-response (MPDR) beamformer. While traditional beamforming relies solely on the microphone signals and thus performs passive noise reduction, recent advancements have proposed integrating a beamformer into an ANC system, i.e., performing noise reduction by jointly processing the microphone signals and playing back anti-noise through the loudspeakers~\cite{Serizel2010,Dalga2011,Patel2020,xiao2023spatial}. These studies have considered two types of target signals, each associated with different delays. The first approach defines the target signal as the delayed desired component in one of the reference microphone signals (or a linear combination)~\cite{Serizel2010,Dalga2011,Patel2020}, where the delay is typically chosen to be half of the filter length to maintain causality. The second approach defines the target signal as the desired component in the error microphone signal without any delay~\cite{xiao2023spatial}. While the first approach is suitable for any fit, the second approach is preferable for open-ear and open-fitting hearables.

In this paper, we systematically investigate the influence of delays for different target signals on the performance of spatially selective ANC for open-fitting hearables. The range of considered delays spans from zero to half of the filter length, which is a common choice for passive noise reduction algorithms. The findings identify an optimal range of delays, particularly when considering the control effort. Moreover, the delay analysis enables us to provide an intuitive explanation for how the system obtains the target signal.

\begin{figure}[t]
    \centering
    \includegraphics[width=\columnwidth]{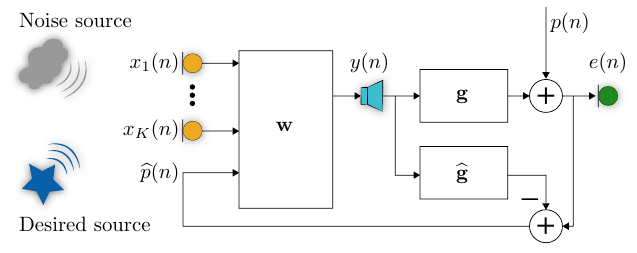}
    \vspace{-2.5em}
    \caption{Block diagram of an ANC system with $K$ reference microphones, one loudspeaker and one error microphone. The control filter is $\mathbf{w}$, the secondary path is denoted by $\mathbf{g}$ and its estimate is denoted by $\widehat{\mathbf{g}}$.}
    \label{fig:1}
    \vspace{-0.5em}
\end{figure}


\section{Signal model}
\label{sec:signal_model}
As shown in Fig.~\ref{fig:1}, we consider an ANC system with $K$ reference microphones. Without loss of generality, we consider one loudspeaker as the secondary source and one error microphone, resulting in a total of $K+1$ microphones. We assume that acoustic feedback from the secondary source to the reference microphones can be perfectly canceled. We assume that the desired sound is speech and is coming from a different direction than the undesired noise. Subscripts $(\cdot)_s$ and $(\cdot)_v$ denote the speech and noise components in signals, respectively.

Before ANC is enabled, the primary sound signal $p(n)$ at the error microphone is given by, 
\begin{equation}
    p(n) = p_s(n) + p_v(n), 
    \label{eq:d_ds_dv}
\end{equation}
where $n$ denotes the time index. 
After ANC is enabled, the error signal is
\begin{equation}
    e(n) = e_s(n) + e_v(n).
    \label{eq:e_es_ev}
\end{equation}

%
The anti-noise at the error microphone can be represented by the product of the stacked input vector ${\mathbf{x}}(n)$ with the stacked control filter ${\mathbf{w}}$ and the convolution matrix $\widetilde{\mathbf{G}}$ of the secondary path, i.e., 
\begin{align}
        e(n) &= p(n) + ( \widetilde{\mathbf{G}}\mathbf{w})^\mathrm{T}{\mathbf{x}}(n) , \label{eq:e_dGwx}
\end{align}
where superscript $(\cdot )^\mathrm{T}$ denotes the transpose.
The stacked control filter ${\mathbf{w}}$ is defined as
\begin{subequations}
\begin{align}
    \mathbf{w} &= \left[\mathbf{w}^\mathrm{T}_1 \ \mathbf{w}^\mathrm{T}_2 \ \dots \ \mathbf{w}^\mathrm{T}_{K+1} \right]^\mathrm{T} \in \mathbb{R}^{(K+1)L_w} ,
    \\
    \mathbf{w}_k &= \left[{w_{k,0}} \ {w_{k,1}} \ \dots \ {w_{k,{L_w}-1} } \right]^\mathrm{T} \in \mathbb{R}^{L_w},  
\end{align}
\end{subequations}
where $L_w$ denotes the control filter length for each channel.
The convolution matrix of the secondary path is given by
\begin{subequations}
\begin{align}
    \widetilde{\mathbf{G}} &= \text{blkdiag}\left({\mathbf{G}} \ {\mathbf{G}} \ \dots \ {\mathbf{G}} \right) ,  \label{eq:G_tilde_multi}
\\ 
    {\mathbf{G}} &= 
        \left[\begin{array}{ccc}
        g_0 & \cdots & 0 \\
        \vdots & \ddots & \vdots \\
        g_{L_g-1} & & g_0 \\
        \vdots & \ddots & \vdots \\
        0 & \cdots & g_{L_g-1}
        \end{array}\right]  \in \mathbb{R}^{(L_g+L_w-1) \times L_w} ,
    \label{eq:G_hat}
\end{align}
\end{subequations}
where $L_g$ is the secondary path filter length.
The stacked input vector ${\mathbf{x}}(n)$ is defined as
\begin{align}
    {\mathbf{x}}(n) &= \left[\mathbf{x}_{1}^\mathrm{T}(n) \ \dots \ \mathbf{x}_{K}^\mathrm{T}(n) \ \widehat{\mathbf{p}}^\mathrm{T}(n) \right]^\mathrm{T} \in \mathbb{R}^{(K+1)(L_g+L_w-1)} , \label{eq:x_tilde_multi}
\end{align}
with
\begin{subequations}
\begin{align}
    \mathbf{x}_k(n) &= \left[ x_k(n) \ \dots \ x_k(n-L_g-L_w+2) \right]^\mathrm{T},
\\ 
    \widehat{\mathbf{p}}(n) &= \left[ \widehat{p}(n) \ \dots \ \widehat{p}(n-L_g-L_w+2) \right]^\mathrm{T},
\end{align}
\end{subequations}
where $\widehat{p}(n)$ is an estimate of the primary sound signal $p(n)$. This estimate can be computed from the error signal $e(n)$ and the secondary source signal $y(n)$ as
\begin{equation}
\widehat{p}(n)=e(n)-\widehat{\mathbf{g}}^\mathrm{T} \mathbf{y}(n),
\label{eq:d_hat}
\end{equation}
where $\widehat{\mathbf{g}}$ denotes an estimate of the secondary path $\mathbf{g} = [ {g}_0\ {g}_1\ \allowbreak \dots\ \allowbreak {g}_{L_g-1} ]^\mathrm{T}$.

Assuming a perfect estimate of the secondary path to be available, i.e., $\widehat{\mathbf{g}} = \mathbf{g}$, such that $\widehat{p}(n) = p(n)$, the primary sound signal can be written as $p(n) = \mathbf{q}^\mathrm{T} \mathbf{x}(n) $, with 
\begin{subequations}
\begin{align}
    \mathbf{q} &= \left[ \mathbf{0}^\mathrm{T} \ \mathbf{0}^\mathrm{T} \ \ldots \ \mathbf{0}^\mathrm{T} \ \bm{\updelta}^\mathrm{T} \right]^\mathrm{T} \in \mathbb{R}^{(K+1)(L_g+L_w-1)} ,    \label{eq:delta_tilde_multi}
\\
    \bm{\updelta} &= \left[1 \;\; 0 \;\; \dots \;\; 0 \right]^\mathrm{T}  \in \mathbb{R}^{(L_g+L_w-1)} ,
\end{align}
\end{subequations}
such that the error signal in~\eqref{eq:e_dGwx} can be written as
\begin{align}
        e(n)=  \mathbf{q}^\mathrm{T}{\mathbf{x}}(n) + (\widetilde{\mathbf{G}}\mathbf{w})^\mathrm{T}{\mathbf{x}}(n) = (\mathbf{q} + \widetilde{\mathbf{G}}\mathbf{w})^\mathrm{T}{\mathbf{x}}(n) .\label{eq:e_qGwx} 
    \end{align}


\section{Spatially selective ANC}
\label{sec:ssanc_delays}
Conventional ANC systems minimize sounds regardless of their incoming directions. A spatially selective ANC system integrates a beamformer into the system such that only undesired sounds from certain directions are minimized, while the desired sound remains. The desired components in the input signals can be separated from the undesired components by computing the relative impulse responses (ReIRs) of the desired source.

\subsection{Cost function and solution}
\label{sec:cost_func_optimal_solution}
We denote the target signal as $t(n)$, which is the signal the system aims to obtain at the error microphone after ANC. There are various possible choices of the target signal in a spatially selective ANC system. The target signal can be defined as a filtered version of the desired component in either one of the reference microphone signals or the error microphone signal.  
Similar to MPDR and LCMP beamformers, while also minimizing the power of the error signal $e(n)$, a spatial constraint based on the ReIRs of the desired source can be applied to the control filter, i.e.,
\begin{equation}
\mathbf{H}^\mathrm{T} (\mathbf{q} + \widetilde{\mathbf{G}}\mathbf{w}) = \mathbf{f},
\label{eq:Hu_f}
\end{equation}
with
\begin{equation}
        \mathbf{H} = \left[ \mathbf{H}_{1} \ \mathbf{H}_{2} \ \dots \ \mathbf{H}_{K+1} \right]^\mathrm{T} \in \mathbb{R}^{(K+1)L \times (L_h+L-1)} , 
        \label{eq:H_bold}
    \end{equation}
where $\mathbf{H}_{k} \in \mathbb{R}^{(L_h+L-1) \times L}$ is the convolution matrix with a similar form as~\eqref{eq:G_hat} of the ReIR $\mathbf{h}_k = [h_{k,0} \ h_{k,1} \ \dots\ h_{k,L_h-1} ]^\mathrm{T}$ between the $k$-th microphone and a chosen \textit{spatial reference microphone} with $L_h$ being the ReIR filter length and $L=L_g+L_w-1$. All ReIRs can be determined from the acoustic impulse responses between the desired source and the microphones. Here, we take the microphone closest to the desired source as the \textit{spatial reference microphone} and assume all ReIRs to be causal. The constraint vector $\mathbf{f} \in \mathbb{R}^{L_h+L-1}$ in~\eqref{eq:Hu_f} reflects the target signal $t(n)$. It can have different definitions depending on different choices of the target signal, which will be discussed in the next subsection.

Using~\eqref{eq:e_qGwx} and~\eqref{eq:Hu_f}, the cost function for a spatially selective ANC system can be defined as
\begin{align}
    \min _{\mathbf{w}} \ & E\left\{e^{2}(n)\right\} + \beta \mathbf{w}^\mathrm{T} \mathbf{w} \notag
    \\
    = \min _{\mathbf{w}} \ & E \left\{(\mathbf{q} + \widetilde{\mathbf{G}}\mathbf{w})^\mathrm{T} \Phi_{{\mathbf{x}} {\mathbf{x}}} (\mathbf{q} + \widetilde{\mathbf{G}}\mathbf{w})\right\} + \beta \mathbf{w}^\mathrm{T} \mathbf{w} \notag
    \\
    \text { s. t. }  & \mathbf{H}^\mathrm{T}(\mathbf{q} + \widetilde{\mathbf{G}}\mathbf{w}) = \mathbf{f}, 
    \label{eq:J_proposed_final}
    \end{align}
where $\Phi_{{\mathbf{x}} {\mathbf{x}}} = E\left\{{\mathbf{x}}(n) {\mathbf{x}}^\mathrm{T}(n)\right\}$ is the autocorrelation matrix of the input vector, with $E\{\cdot\}$ being the mathematical expectation operator, and $\beta$ being a control effort weighting factor.

The solution of~\eqref{eq:J_proposed_final} is given by~\cite{xiao2023spatial}
\begin{align}
\mathbf{w} =  
&- \left[ \mathbf{I} - \Phi_{{\mathbf{r}} {\mathbf{r}}}^{-1} \widetilde{\mathbf{G}}^\mathrm{T} \mathbf{H} (\mathbf{H}^\mathrm{T} \widetilde{\mathbf{G}} \Phi_{{\mathbf{r}} {\mathbf{r}}}^{-1} \widetilde{\mathbf{G}}^\mathrm{T} \mathbf{H} + \rho \mathbf{I} )^{-1} \mathbf{H}^\mathrm{T} \widetilde{\mathbf{G}} \right] \Phi_{{\mathbf{r}} {\mathbf{r}}}^{-1} \phi  \notag
   \\
&+ \Phi_{{\mathbf{r}} {\mathbf{r}}}^{-1} \widetilde{\mathbf{G}}^\mathrm{T} \mathbf{H} ( \mathbf{H}^\mathrm{T} \widetilde{\mathbf{G}} \Phi_{{\mathbf{r}} {\mathbf{r}}}^{-1} \widetilde{\mathbf{G}}^\mathrm{T} \mathbf{H} + \rho \mathbf{I} )^{-1} \left( \mathbf{f} - \mathbf{H}^\mathrm{T} \mathbf{q} \right), 
\label{eq:w_opt}
\end{align}
\mathcenter
with 
\begin{equation}
    \Phi_{{\mathbf{r}} {\mathbf{r}}} = \widetilde{\mathbf{G}}^\mathrm{T} \Phi_{{\mathbf{x}} {\mathbf{x}}} \widetilde{\mathbf{G}} + \beta \mathbf{I}, \qquad \quad \phi = \widetilde{\mathbf{G}}^\mathrm{T} \Phi_{{\mathbf{x}} {\mathbf{x}}} \mathbf{q}  ,
\end{equation}
where $\mathbf{I}$ denotes the identity matrix, and $\rho$ is a regularization factor due to matrix $\mathbf{H}^\mathrm{T} \widetilde{\mathbf{G}} \Phi_{{\mathbf{r}} {\mathbf{r}}}^{-1} \widetilde{\mathbf{G}}^\mathrm{T} \mathbf{H}$ possibly being rank-deficient (e.g., due to delays in the secondary path)~\cite{hansen1998rank}.

\subsection{Target signal and delays}
\label{sec:target_signal_delays}
Studies~\cite{Serizel2010,Dalga2011,Patel2020,xiao2023spatial} all similarly minimized the noise component. However, they had different target signals to obtain.

In~\cite{Serizel2010,Dalga2011,Patel2020}, the target signal was defined as the delayed desired component at a reference microphone (e.g., the \textit{spatial reference microphone}), that is, $t(n)=x_{\text{ref},s}(n-\Delta)$. In this case, vector $\mathbf{f} $ is given by
\begin{equation}
\mathbf{f} = \bm{\Uppsi} \bm{\updelta}_\Delta, \quad \bm{\updelta}_\Delta = [ \underbrace{0 \ \dots \ 0}_{\Delta} \ 1 \ \ 0 \ \dots \ 0]^\mathrm{T} \in \mathbb{R}^{L_h},
\label{eq:Hu_f_delayed_ref}
\end{equation}
where the spectral weighting matrix $\bm{\Uppsi} \in \mathbb{R}^{(L_h+L-1) \times L_h}$ is the convolution matrix from a minimum-phase high-pass filter $\bm{\uppsi} \in \mathbb{R}^{L}$ (e.g., cut off at 120~Hz). Such a spectral weighting method can be used to improve the noise reduction performance at the cost of some signal distortion~\cite{xiao2023spatial}. 

In~\cite{xiao2023spatial}, the target signal was defined as the desired component at the {error} microphone, which may also include certain delays, i.e., $t(n)=p_s(n-\Delta)$. In this case, vector $\mathbf{f} $ is given by
\begin{align}
    \mathbf{f} &= \bm{\Uppsi} \mathbf{h}_{K+1,\Delta} , 
    \label{eq:f}
    \end{align}
where $\mathbf{h}_{K+1,\Delta}$ is the ReIR from the spatial reference microphone to the error microphone $\mathbf{h}_{K+1}$ delayed by $\Delta$ samples. 

\section{Simulations}
\label{sec:simulations}
In this section, we systematically evaluate the influence of delays on the system performance. Section~\ref{sec:setup_metrics} discusses the acoustic setup, algorithm parameters and evaluation metrics. Sections~\ref{sec:speech_at_error} and~\ref{sec:speech_at_ref} present simulation results when defining the target signal either at the error microphone or at the spatial reference microphone.

\subsection{Setup and evaluation metrics} \label{sec:setup_metrics}    
\begin{figure}[t]
    \centering
    \includegraphics[width=0.99\columnwidth]{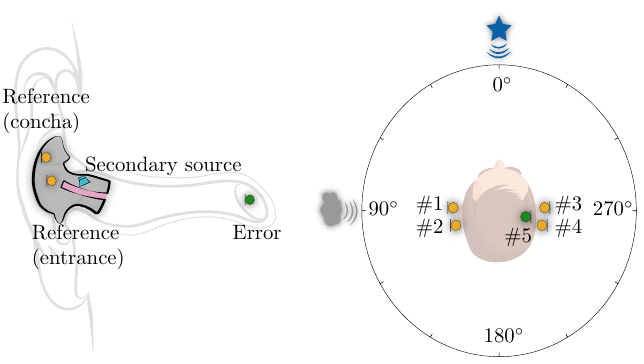}
    \vspace{-1.5em}
    \caption{Illustration of the open-fitting hearable, the considered microphones and the simulation setup. }
    \vspace{-0.5em}
    \label{fig:2}
\end{figure}

For the simulations, we considered a pair of open-fitting hearables~\cite{denk2019one,denk2021hearpiece} inserted in both ears of a GRAS 45BB-12 KEMAR Head \& Torso simulator, as shown in Fig.~\ref{fig:2}. We used four reference microphones (concha microphones at the left and right ears, entrance microphones at the left and right ears, labeled as \#1 -- \#4), one error microphone (located at the right ear, labeled as \#5) and one secondary source (outer receiver at the right ear). The error microphone was assumed to be at the eardrum. To generate the microphone signals and compute the ReIRs, we used the database from~\cite{denk2021hearpiece}, which contains measured impulse responses in an anechoic chamber for a source at various directions relative to these microphones. In this setup, we considered a desired clean speech source from $0^\circ$ (\say{p234\_005} from the VCTK dataset~\cite{veaux2017cstr}) and a noise source from $90^\circ$ (babble noise from the NOISEX-92 database~\cite{Varga1993}). The signals had a duration of 5~s with a sampling rate of 16~kHz. The signal-to-noise ratio (SNR) at the error microphone was set to --5~dB. The filter lengths of the control filter, the secondary path and the ReIRs were equal to $L_w=L_g=L_h=280$. The causal ReIRs were computed using the least-mean-squares adaptive filter (after convergence) from microphone signals simulated with white noise at $0^\circ$, using the entrance microphone \#3 as the spatial reference microphone. The high-pass filter $\bm{\uppsi}$ had a cut-off frequency at 120~Hz. In all cases, $\beta = \lambda_\text{max} (\widetilde{\mathbf{G}}^\mathrm{T} \Phi_{{\mathbf{x}} {\mathbf{x}}} \widetilde{\mathbf{G}} ) / 500$ and $\rho = \lambda_\text{max} (\mathbf{H}^\mathrm{T} \widetilde{\mathbf{G}} \Phi_{{\mathbf{r}} {\mathbf{r}}}^{-1} \widetilde{\mathbf{G}}^\mathrm{T} \mathbf{H} ) / 30000$, where $\lambda_\text{max}(\cdot )$ denotes the largest eigenvalue.

The following four metrics are used for evaluation. The noise reduction (NR) level of the ANC is defined as
\begin{equation}
    \text{NR (dB)} = 10\log_{10}\left( \frac{ \sum\limits_{n=1}^{N}  p_v^2(n)  }{ \sum\limits_{n=1}^{N} e_v^2(n) } \right),
\end{equation}
where $e_v(n)$ can be obtained by using $\mathbf{w}$ in~\eqref{eq:w_opt} to filter only the noise components in the signals. $N$ denotes the signal length.

The speech distortion index (SDI) is used to assess distortion in the speech component at the error microphone after ANC~\cite{Chen2006}. It is defined as
\begin{equation}
    \text{SDI (dB)} = 10\log_{10} \left( \frac{ \sum\limits_{n=1}^{N} [t(n) - e_s(n)]^2 }{ \sum\limits_{n=1}^{N} t^2(n) } \right),
\end{equation}
where $e_s(n)$ can be obtained by using $\mathbf{w}$ to filter only the speech components in the signals.

To assess the overall signal quality, we used the narrowband perceptual evaluation of speech quality (NB-PESQ) mean opinion score - listening quality objective (MOS-LQO)~\cite{Rix2001PESQ}, either between $t(n)$ and $p(n)$ (ANC off) or between $t(n)$ and $e(n)$ (ANC on). It should be noted that the reference signals in the two cases are different, since the definitions of the target signal are different.

Finally, we computed the control effort, an important factor to consider in ANC systems~\cite{Elliott2000}, which is defined as
\begin{equation}
    \mathcal{E} = \sum_{n=1}^{N} y^2(n) .
\end{equation}

\subsection{Target signal at the error microphone}
\label{sec:speech_at_error}
\begin{figure}[t]
    \centering
    \captionsetup{aboveskip=0.1\normalbaselineskip}
    \includegraphics[width=0.94\columnwidth]{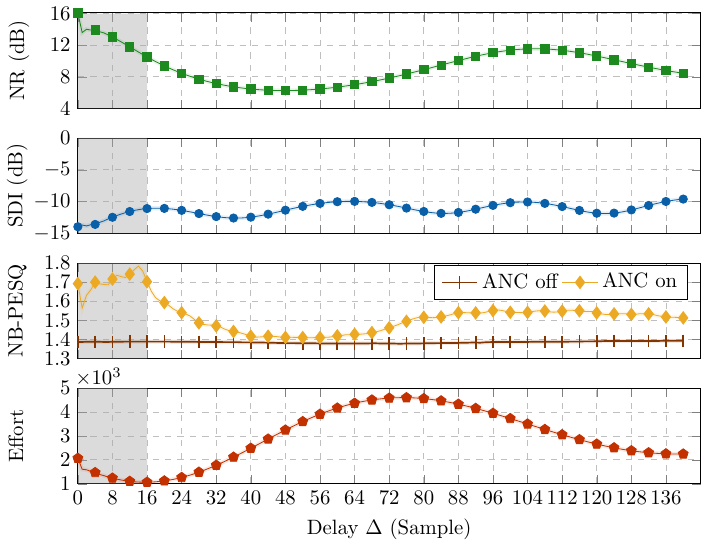}
    \vspace{-1em}
    \caption{The NR level, SDI, NB-PESQ MOS-LQO score and control effort for $\Delta \in [0:1:140]$ sample delays in the target signal (desired speech component at the \textit{error} microphone). The shaded areas indicate the recommended range for the delay.}
    \label{fig:3}
\end{figure}

Figure~\ref{fig:3} depicts the evaluation metrics for various delays, ranging from 0 to 140 (half of the control filter length, $L_w/2$) with a step size of 1, when the target signal is defined as the delayed speech component at the error microphone, i.e., $t(n)=p_s(n-\Delta)$.

For all considered metrics, it can be observed that the best performance is obtained for small delays. In fact, for all metrics except control effort, the best performance is obtained for $\Delta=0$. The lowest value for the control effort is obtained for $\Delta=16$, but for this value, the NR level is also degraded. For example, the NR level is 16.1~dB for $\Delta=0$, but 10.5~dB for $\Delta=16$, although the control effort decreased from 2065 to 1063.
\begin{figure}[!ht]
    \centering
    \vspace{-1em}
    \includegraphics[width=0.92\columnwidth]{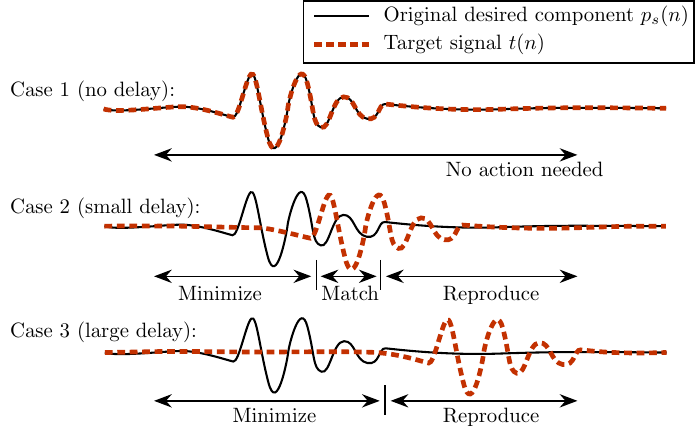}
    \caption{Illustration of the mechanism behind selective ANC for various delays between the original desired component at the error microphone $p_s(n)$ and the target signal $t(n)$.}
    \label{fig:4}
\end{figure}

For three cases (no delay, small delay, large delay), Fig.~\ref{fig:4} visualizes the mechanism behind selective ANC, depicting the original speech component in the error microphone, $p_s(n)$, and the target signal, $t(n)=p_s(n-\Delta)$. For Case 1 (no delay), the original speech component just needs to be preserved, requiring no action for this system~\cite{xiao2023spatial}. For Case 2 (small delay), there are three mechanisms involved: minimizing the original speech component, matching the original speech component to the target, and reproducing the target signal. This will typically require more control effort than for $\Delta=0$. The extreme case is shown in Case~3 (large delay), where the original speech component and the target are completely misaligned due to the delay. The system will need to minimize the original speech component first and then reproduce the target once again, thus requiring the most effort. This may explain why for $\Delta \geq 40$, there is a significant increase of the control effort but with low NR levels.

In summary, these simulation results indicate that although a slight delay is permissible ($\Delta \leq 16$, as seen in the shaded areas in Fig.~\ref{fig:3}), it is still best to design the delay equal to zero when the target is defined at the error microphone.

\subsection{Target signal at the spatial reference microphone}
\label{sec:speech_at_ref}
\begin{figure}[t]
    \centering
    \captionsetup{aboveskip=0.1\normalbaselineskip}
    \includegraphics[width=0.94\columnwidth]{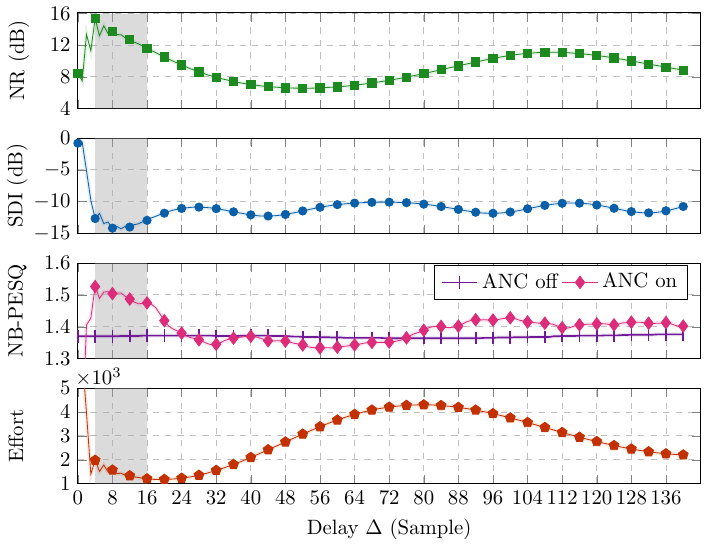}
    \vspace{-1em}
    \caption{The NR level, SDI, NB-PESQ MOS-LQO score and control effort for $\Delta \in [0:1:140]$ sample delays in the target signal (desired speech component at the \textit{spatial reference} microphone). The shaded areas indicate the recommended range for the delay.}
    \vspace{-1em}
    \label{fig:5}
\end{figure}

Figure~\ref{fig:5} depicts the evaluation metrics for the same various delays, when the target signal is defined as the delayed speech component in the spatial reference microphone, i.e., $t(n)=x_{\text{ref},s}(n-\Delta)$. For all considered metrics, it can be observed that the worst performance is obtained for $\Delta=0$.
Since the system tries to obtain the original speech component from the spatial reference microphone at the error microphone, causality cannot be satisfied~\cite{Serizel2010,ranjan2015natural,schepker2022robust}, thus requiring certain delays.

In~\cite{Serizel2010,Dalga2011}, a delay of half of the filter length was used. For our simulations, using $\Delta=140$ results in 8.9~dB NR, $-10.8$~dB SDI, 0.03 PESQ improvement and a control effort of 2207. However, as depicted in the shaded areas of Fig.~\ref{fig:5}, it is preferable to apply only a minor delay ($4 \leq \Delta \leq 16$) to achieve the largest NR and PESQ improvement, as well as the lowest distortion and control effort. For example, $\Delta=4$ results in 15.4~dB NR level, --12.7~dB SDI, 0.16 PESQ improvement with a control effort of 1983. In fact, the acoustic delay between the spatial reference microphone and the error microphone from the desired source also has four samples. Therefore, it is sensible to allow for a four-sample delay in the system.

In summary, these simulation results indicate that when the target signal is defined in a reference microphone, excessive delays result in inefficient control and small (non-zero) delays are preferred instead. The mechanism can be similarly explained as in Fig.~\ref{fig:4}.

\section{Conclusion}
\label{sec:conclusion}
This paper examined two types of target signals with various delays in a pair of spatially selective open-fitting ANC hearables. When the target signal is the desired component in the error microphone signal, optimal performance was achieved when the original desired component was preserved without any delay. However, when the target signal is the desired component in a reference microphone signal, a small delay is required to satisfy causality. The optimal delay is found to be the acoustic delay between this reference microphone and the error microphone from the desired source. Using large delays leads to degraded performance and increased control effort as the system attempts to minimize the original speech component before reproducing the delayed version.

\section{References}

%
\printbibliography[heading=none]

\end{document}